\documentclass[11pt]{article}
\pdfoutput=1 
\usepackage{jheppub}
\usepackage[utf8]{inputenc}

\usepackage{graphicx}
\usepackage{color, graphics}
\usepackage{amsmath,amssymb,amsfonts}
\usepackage{verbatim}   
\usepackage{subfig}  
\usepackage{hyperref}
\usepackage{mathrsfs}
\usepackage{multirow}
\usepackage{url}
\usepackage{slashed,afterpage}
\usepackage{hyperref}

\newcommand{\MeV}{{\, {\rm MeV}}}
\newcommand{\GeV}{{\, {\rm GeV}}}
\newcommand{\TeV}{{\, {\rm TeV}}}

\newcommand{\der}{\mathrm{d}}

\def\beq{\begin{equation}}
\def\eeq{\end{equation}}
\def\bea{\begin{eqnarray}}
\def\eea{\end{eqnarray}}
\def\bitem{\begin{itemize}}
\def\eitem{\end{itemize}}
\newcommand{\bec}{\begin{center}}
\newcommand{\eec}{\end{center}}
\newcommand{\ba}{\begin{array}}
\newcommand{\ea}{\end{array}}

\def\imp{~\Rightarrow}

\let\<=\langle
\let\>=\rangle

\def\OO{\mathcal{O}}

\begin{document}

\title{Glueball dark matter in non-standard cosmologies}
\author[a,b]{Bobby S. Acharya,}
\author[a]{Malcolm Fairbairn,}
\author[b,c]{Edward Hardy}
\emailAdd{bobby.acharya@kcl.ac.uk}
\emailAdd{malcolm.fairbairn@kcl.ac.uk}
\emailAdd{ehardy@ictp.it}
\affiliation[a]{Kings College London, Strand, London, WC2R 2LS, United Kingdom}
\affiliation[b]{Abdus Salam International Centre for Theoretical Physics, Strada Costiera 11, 34151, Trieste, Italy}
\affiliation[c]{Department of Mathematical Sciences, University of Liverpool, Liverpool L69 7ZL, United Kingdom}

\abstract{
Hidden sector glueball dark matter is well motivated by string theory, compactifications of which often have extra gauge groups uncoupled to the visible sector. 
We study the dynamics of glueballs in theories with a period of late time primordial matter domination followed by a low final reheating temperature due to a gravitationally coupled modulus. 
Compared to scenarios with a high reheating temperature, the required relic abundance is possible with higher hidden sector confinement scales, and less extreme differences in the entropy densities of the hidden and visible sectors. Both of these can occur in string derived models, and  relatively light moduli are helpful for obtaining viable phenomenology. We also study the effects of hidden sector gluinos. In some parts of parameter space these can be the dominant dark matter component, while in others their abundance is much smaller than that of glueballs. Finally, we show that heavy glueballs produced from energy in the hidden sector prior to matter domination can have the correct relic abundance if they are sufficiently long lived.

}

\maketitle

\section{Introduction}

Pure gauge dark sectors are common in string theory UV completions of the Standard Model (SM), and if they have no significant couplings to the visible sector typically contain  cosmologically stable glueballs \cite{Okun:1980mu,Faraggi:2000pv}.\footnote{By gluons/ glueballs, we always mean dark sector gluons/ glueballs unless explicitly stated.} These can be viable dark matter (DM) candidates, potentially with self interactions of astrophysically relevant strength, which might resolve disagreement between cold dark matter simulations and observations of small scale structure \cite{Spergel:1999mh}.

Previous work on glueball DM has assumed a high reheating temperature, above the scale of the confining phase transition in the hidden sector. The dynamics of such models are equivalent to cannibal dark matter theories \cite{Carlson:1992fn} with the final glueball abundance set by a combination of the thermal abundance of dark gluons at the time of the hidden sector phase transition and subsequent $3\rightarrow 2$ interactions.  
In typical models, obtaining a sufficiently small glueball relic abundance requires an extremely large initial ratio of entropies in the visible and hidden sectors. From the perspective of underlying theories in which the SM is just one, undistinguished, sector among many this is somewhat troubling.\footnote{One possibility for accommodating such sectors is if the universe is reheated last by decays of the inflaton, which might couple to hidden sector gauge fields only very weakly.}

However in many compactifications of string theory, especially with low scale supersymmetry motivated by naturalness and gauge coupling unification, there are  slow decaying light moduli \cite{Coughlan:1983ci,Goncharov:1984qm,Ellis:1986zt,Banks:1993en,deCarlos:1993wie,Acharya:2008bk,Acharya:2010af}. At late times, the longest lived of these will dominate the energy density of the universe, leading to an extended period of matter domination, and depleting other energy. The eventual decay of the last modulus repopulates the visible and hidden sectors with radiation, and the final reheat temperature is fairly low. The lightest mass eigenstate is often a linear combination of multiple moduli, and it is reasonable to expect that all sectors are reheated significantly.

In this paper we study how a low reheat temperature changes the constraints on pure gauge hidden sectors and glueball dark matter. Provided the modulus mass is above the scale of the hidden sector phase transition $\Lambda$, glueballs can still be produced during reheating. However, if the energy density transfered to the hidden sector is $\lesssim \Lambda^4$, the subsequent dynamics are modified compared to the high reheat temperature case. A glueball relic abundance consistent with the observed DM value is possible with significantly larger hidden sector confinement scales, and relatively mild hierarchies between the initial visible and hidden sector energy densities. 

Given the connection with string theory it is also interesting to study how this maps onto constraints on the gauge groups and couplings at a high scale, and the consequences for UV model building. Pure gauge sectors with couplings at $10^{16}~\GeV$ comparable to the unified value of the visible sector couplings have high confinement scales \cite{Halverson:2016nfq}. From this perspective our motivating argument can be reversed-- considering an underlying model with multiple pure gauge hidden sectors not coupled to the visible sector, and the expectation that the inflaton and moduli do not couple very differently to separate sectors, 
\emph{light moduli with masses $\sim {\it 10^{5}}~GeV$ are useful to avoid unacceptable phenomenology}. This is further motivated since the lightest modulus cannot decay to sectors with a confinement scale above its mass, so does not lead to a large relic density in heavy glueballs.

In addition to the scenario presented above, we investigate two related scenarios which might arise in the same setup.  First we consider the situation where supersymmetry is broken  weakly in the hidden sector such that there may be dark gluinos with masses below that of the lightest modulus. In some models with a non-thermal cosmology these are the dominant dark matter component, while in others they are negligible compared to glueballs. Second, we return to the glueball as dark matter but taking into account different initial conditions - while over most of parameter space any energy in the glueball sector before matter domination is completely irrelevant afterwards, this is not always the case. If the hidden sector confinement scale is high, glueballs produced during the matter dominated era can lead to the correct relic density, allowing heavy DM with mass $\sim 10^{9}~\GeV$. Such models do however require that the glueballs have extremely weak couplings to all lighter moduli, otherwise they quickly decay and are 
cosmologically irrelevant. 

The structure of this paper is as follows: in Section~\ref{sec:high} we collect and review results on glueball DM in models with a high reheat temperature and a standard cosmological history. In Section~\ref{sec:low} we study how a non-thermal cosmology changes the viable parameter space. 
In Section~\ref{sec:gluino} we study the possible role of hidden sector gluinos, and in Section~\ref{sec:resid} we consider scenarios in which a significant population of glueballs forms during matter domination. Finally in Section~\ref{sec:discuss} we discuss our results.

\section{Glueball dark matter in the standard cosmology} \label{sec:high}

If the hidden sector reheat temperature is significantly above the scale of the confining phase transition, the hidden sector gluons quickly reach a thermalised distribution with number density
\beq
n_g \simeq \frac{\xi\left(3\right)}{\pi^2} g_g T_{g}^3  ~,
\eeq
where $T_g$ is the glueball sector temperature,  and  $g_g$ the effective number of degrees of freedom in the glueball sector. We define
\beq \label{eq:Bdefhigh}
B \equiv \frac{g_v \rho_g}{g_g \rho_v} = \frac{T_g^4}{T_v^4} ~,
\eeq
evaluated at an early time, but once both sectors have reached chemical equilibrium, where $g_v$ is the visible sector effective number of degrees of freedom, $\rho_v$ and $\rho_g$ the visible and glueball sector energy densities, and $T_v$ the visible sector temperature. This 
is motivated by initial conditions set by perturbative inflaton decays, with $B$ approximately measuring the deviation from universal couplings.\footnote{However, $g_v$ still includes the usual $7/8$ correction for fermions.} The corresponding ratio of entropies is
\beq
\frac{s_{g}}{s_v} = \frac{g_g}{g_v} B^{3/4}~.
\eeq
For viable models $B \ll 1$, and the expansion of the universe is set by the visible sector. Provided there is no subsequent entropy injection, if there are no significant couplings between the visible and hidden sectors entropy is separately conserved in each.\footnote{Inflaton mediated thermalisation can be important in some models \cite{Adshead:2016xxj,Hardy:2017wkr}.}

When the dark sector temperature drops below the scale of its confining phase transition, each gluon will result in an order 1 glueball being produced. For simplicity we assume exactly one glueball is produced per gluon, so do not distinguish between, gluon and glueball number densities. ${\rm SU}\left(N\right)$ gauge theories have complicated phase transitions, which are strongly first order in the large $N$ limit \cite{Lucini:2003zr,Lucini:2005vg,Garcia:2015loa}. The effect of this on the eventual glueball relic abundance has been studied in \cite{Forestell:2016qhc}, but only leads to $\OO\left(1\right)$ changes to the parameter space, so is neglected here. Additionally, there is likely to be a spectrum of glueball states \cite{Morningstar:1999rf}, which can also have an $\OO\left(1\right)$ impact on the glueball dynamics and abundance \cite{Forestell:2016qhc}.

The glueball number density immediately after confinement is equivalent to a yield $y_g = n_g/s_v$ of
\beq \label{eq:yieldnrh}
y_g \simeq \frac{45 \xi\left(3\right)}{2\pi^4} \frac{g_g}{g_v} \left(\frac{T_{g}}{T_{v}}\right)^{3} \simeq \frac{45 \xi\left(3\right)}{2\pi^4} \frac{g_g}{g_v}  B^{3/4}~.
\eeq
This is a reasonable approximation to the final glueball yield, and we note that the correct relic density requires very small values of $B$, roughly 
\beq \label{eq:bapprox}
B \simeq  3\times 10^{-10} \frac{1}{\left(N^2-1\right)^{4/3}} \left(\frac{\GeV}{\Lambda}\right)^{4/3} ~,
\eeq
assuming $g_v$ has the high temperature SM value, and that the glueball sector is an ${\rm SU}\left(N\right)$ gauge theory.

The glueballs have number changing interactions, and in large parts of parameter space these are efficient on the timescale of Hubble expansion immediately after the confining phase transition \cite{Boddy:2014yra,Forestell:2016qhc}. 
Subsequently, as their temperature decreases, the glueballs become non-relativistic and their equilibrium number density $n_{\rm eq}$ drops as 
\beq \label{eq:neq}
n_{\rm eq} =  \left(\frac{\Lambda T_{g}}{2\pi}\right)^{3/2} e^{-\Lambda/T_g} ~,
\eeq
assuming the number of glueball degrees of freedom equals $1$ below the confinement scale. The number density of glueballs initially stays close to the equilibrium value, dominantly through $3 \rightarrow 2$ processes, which has the effect of heating up the dark sector relative to the visible sector \cite{Carlson:1992fn}. Other DM models involving similar dynamics include  \cite{Hochberg:2014dra,Yamanaka:2014pva,Bernal:2015ova,Bernal:2015xba,Kuflik:2015isi,Soni:2016gzf,Pappadopulo:2016pkp,Farina:2016llk}. The cross section for $3 \rightarrow 2$ glueball interactions for an ${\rm SU}\left(N\right)$ gauge theory in the large $N$ limit is expected to scale as
\beq \label{eq:32scal}
\left<\sigma v^2\right>_{3\rightarrow 2} \simeq \frac{\left(4\pi \right)^3}{ N^6 \Lambda^5} ~,
\eeq
up to an order 1 constant \cite{Forestell:2016qhc}. Since entropy is conserved, once the glueballs are non-relativistic their number density and temperature are related to the visible sector entropy by
\beq \label{eq:snr}
s_{v} = \frac{g_v s_g}{g_g B^{3/4}} \simeq \frac{g_v}{g_g B^{3/4}} \frac{\Lambda n_g}{T_g} ~,
\eeq
where $g_g$ is still the number of degrees of freedom in the gluon sector above the confinement scale.

At later times $3 \rightarrow 2$ processes freeze out, and the glueball number density stops tracking the equilibrium value, Eq.~\eqref{eq:neq}. Decoupling happens when interactions can no longer reduce the glueball number density as fast as the equilibrium number density drops due to the expansion of the universe, that is when $\Gamma_{3\rightarrow 2} \lesssim \dot{n}_{e}/n_{\rm eq}$. This is equivalent to
\beq \label{eq:de32}
\Gamma_{3\rightarrow 2} = \frac{3 T_{\rm FO}}{\Lambda} H_{\rm FO} ~,
\eeq
where $H_{\rm FO}$ is the Hubble parameter at decoupling, $T_{\rm FO}$ is the glueball temperature at this time, and $\Gamma_{3\rightarrow 2} = n_g^2 \left<\sigma v^2\right>_{3\rightarrow 2}$ the rate of $3\rightarrow 2$ processes. In the part of parameter space where freeze out happens for a temperature $\Lambda/T_g \ll 1$, using the non-relativistic formula for entropy Eq.~\eqref{eq:snr} is valid. Combining Eqs.~\eqref{eq:neq},\eqref{eq:snr}, and \eqref{eq:de32}, the glueball temperature at freeze out is
\beq
T_{\rm FO} = \frac{4 \Lambda}{5}~ W\!\left[0.064~ g_g^{2/5} g_v^{-3/10} B^{3/10} M_{\rm Pl}^{3/5} \Lambda^{12/5} \left<\sigma v^2\right>_{3\rightarrow 2}^{3/5}  \right]^{-1} ~,
\eeq
where $W\!\left[x\right]$ is the product-logarithm (that is, $W\!\left[x\right]$ is the inverse of $x e^x$). After $3\rightarrow 2$ interactions stop, $2\rightarrow 2$ scattering between glueballs remains efficient. However $2 \rightarrow n$ processes (with $n>2$) are negligible because the glueballs are non-relativistic and there is little kinetic energy available for production of extra states.  As a result the complicated details of the $2 \rightarrow n$ interactions are not important.  At early times they happen fast but simply keep the system in equilibrium,  meanwhile at late times after $3 \rightarrow 2$ freeze out they only change the relic abundance slightly.

To match the present day observed DM abundance requires a final yield 
\beq
y_{\infty}  \simeq \frac{4.4\times 10^{-10}~\GeV}{\Lambda}~ ,
\eeq
and using
\beq
y_{\infty} = \frac{g_g T_{\rm FO}}{g_v \Lambda} B^{3/4} ~,
\eeq
the glueball relic abundance $\left(\Omega h^2\right)_{G}$ is
\beq
\frac{\left(\Omega h^2\right)_{G}}{\left(\Omega h^2\right)_{\rm DM}}  = 0.056 \left(N^2-1 \right)  \left(\frac{B}{10^{-12}}\right)^{3/4} \left(\frac{\Lambda}{\GeV}\right) ~W\!\left[2.1\frac{\left(N^2-1\right)^{2/5}}{N^{18/5}} B^{3/10} \left(\frac{M_{\rm Pl}}{\Lambda}\right)^{3/5} \right]^{-1} ~,
\eeq
for an ${\rm SU}\left(N\right)$ sector with $g_v$ equal to the SM value, where $\left(\Omega h^2\right)_{\rm DM}$ is the observed dark matter value. 
This expression is valid if the glueballs become non-relativistic before $3\rightarrow 2$ process decouple. If $3\rightarrow 2$ processes freeze out before the glueballs become non-relativistic, there will be no significant period of time when the yield is reduced by tracking the non-relativistic equilibrium number density. Instead it is given by Eq.~\eqref{eq:yieldnrh}, up to a correction of $\lesssim 2$, since $2\rightarrow n$ processes could convert some of the remaining kinetic energy to extra glueballs.

Glueball DM can have self-interactions at late times, potentially with observable astrophysical consequences.\footnote{As well as glueball dark matter, there are many other possible hidden sector dark matter scenarios, many of which can lead to self-interactions, for example \cite{Kribs:2009fy,Buen-Abad:2015ova,Gross:2015cwa,Eby:2015hsq,Kribs:2016cew,Dienes:2016vei}.} These could alleviate tensions between numerical simulations of cold dark matter and small scale structure observations  \cite{Faraggi:2000pv,Boddy:2014yra}. However, significant uncertainty still remains in simulations, for example due to baryons not being included, and it is plausible that future developments will lead to agreement without self-interactions. Further details and discussion may be found in, for example, \cite{Spergel:1999mh,deBlok:2009sp,Rocha:2012jg,Peter:2012jh,Weinberg:2013aya,Tulin:2013teo,Cline:2013zca,Elbert:2014bma,Harvey:2015hha}. Since the strength of the interactions are controlled solely by the strong coupling scale $\Lambda$, and are independent of the details of the early universe, we simply note that values of $\Lambda \simeq 100~\MeV$ lead to astrophysically relevant self-interactions. Meanwhile models with significantly smaller $\Lambda$ are excluded, while for $\Lambda \gg 100~\MeV$ late time self-interactions are not important.

The glueball phase transition can be strongly first order and produce gravitational waves \cite{Schwaller:2015tja}. Unfortunately their intensity is suppressed by $\sim \left(\rho_g / \rho_v \right)^{2}$ if they are dominantly from bubble collisions and $\sim \left(\rho_g / \rho_v \right)^{3/2}$ if hydrodynamical turbulence leads to significant emission \cite{Caprini:2009fx,Huber:2008hg,Caprini:2009yp}. For values of $ \left(\rho_g / \rho_v \right)$ compatible with the DM relic abundance, the suppression is sufficiently large that observation is not possible in currently planned experiments, at least for the minimal models consider.

\begin{figure}
	\begin{center}
		\includegraphics[width=.6\textwidth]{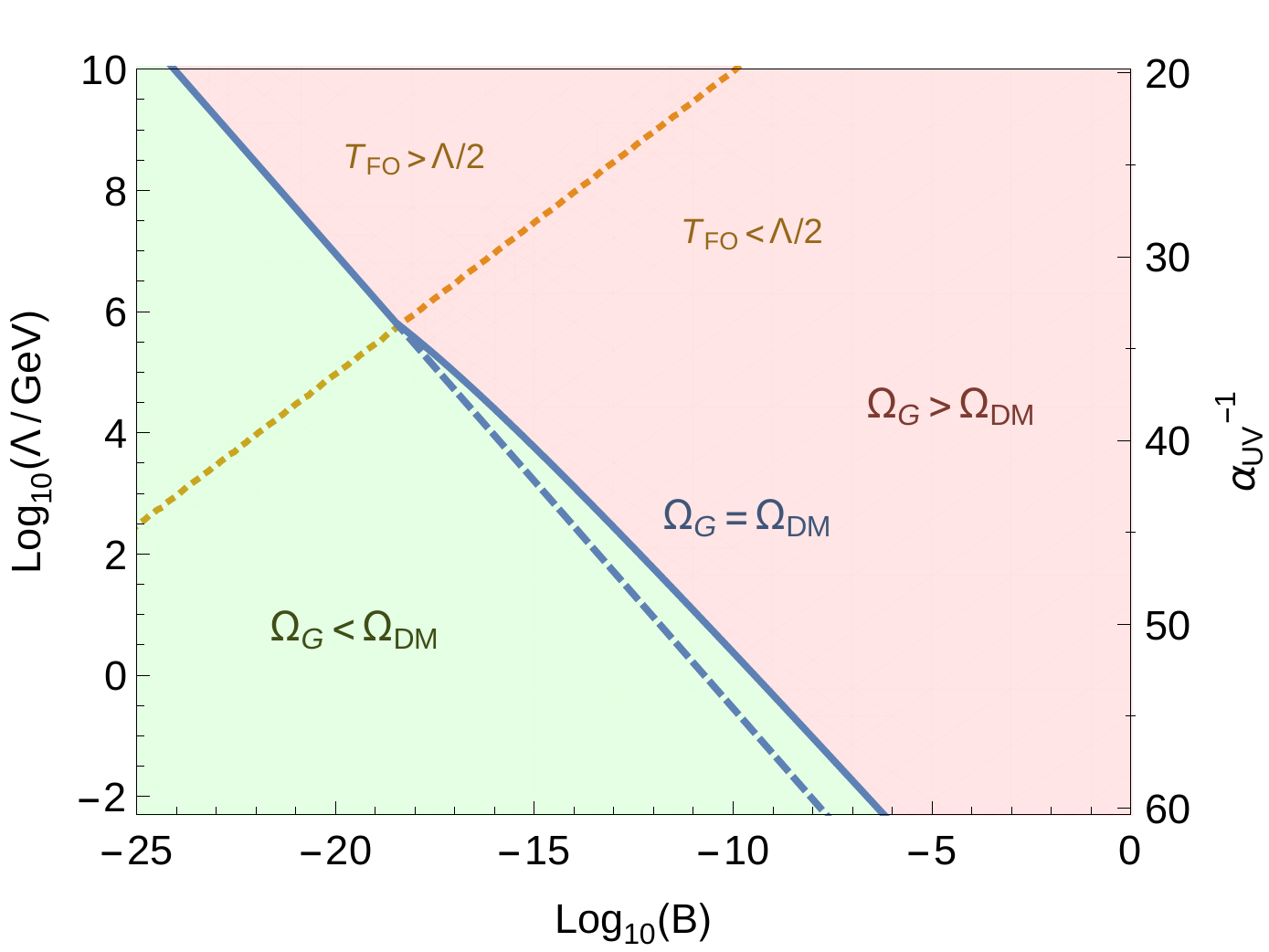}
	\end{center}
	\caption{
		The viable glueball DM parameter space for an ${\rm SU}\left(3\right)$ hidden sector in a model with a high reheat temperature and standard cosmology. $B$ parameterises the relative strength of inflaton decays to the glueball and visible sectors, defined in Eq.~\eqref{eq:Bdefhigh}, and $\Lambda$ is the hidden sector confinement scale. Green regions have an under abundance of glueball dark matter, and models on the solid blue line have the observed DM relic density. 
		 The results obtained ignoring $3\rightarrow 2$ interactions are shown in dashed blue. Below $\Lambda \simeq 0.1~\GeV$ late time DM self interactions are significant and such models are probably ruled out. Above and left of the orange line  $3\rightarrow 2$ interactions freeze out when the glueball temperature is $> \frac{1}{2} \Lambda$. The UV values of the hidden sector gauge couplings $\alpha_{\rm UV}$ defined at $10^{16}~\GeV$ corresponding to $\Lambda$ are also shown, assuming the theory is supersymmetric above $10^6~\GeV$.
	}
	\label{fig:1}
\end{figure}

The viable parameter space for an ${\rm SU}\left(3\right)$ theory is shown in Fig.~\ref{fig:1}, assuming the visible sector effective number of degrees of freedom is that of the Standard Model. It is also well motivated to take the Minimal Supersymmetric Standard Model value at scales above $\sim \TeV$, which would make an $\OO\left(1\right)$ difference to the results. To get the observed DM relic abundance, or an under-abundance,  a large initial entropy ratio is needed for all allowed values of $\Lambda$. We also show the constraints that would be obtained ignoring the late time $3 \rightarrow 2$ processes, which make a significant, but not dramatic, difference. For very large $\Lambda$, close to the top of the plot, only models with a heavy inflaton that decays relatively fast will reheat the glueball sector above its strong coupling scale. If this is not the case the dynamics are instead as studied in Section~\ref{sec:low}.

The results can also be recast as constraints on high scale physics. For a supersymmetric ${\rm SU}\left(N\right)$ pure gauge sector, the beta function is
\beq
\frac{\der \alpha}{\der \log \mu} = -\frac{3 N}{2\pi} \alpha^2~,
\eeq
where $\mu$ is the renormalisation scale, and a theory with gauge coupling ${\alpha}_{\rm UV}$ at a UV scale $\Lambda_{\rm UV}$ has a strong coupling scale
\beq \label{eq:run}
\Lambda = \Lambda_{\rm UV}\exp\left(-\frac{2\pi}{3 N} \frac{1}{ \alpha_{\rm UV}} \right)~.
\eeq
It is straightforward to transform the allowed parameter space in terms of $\Lambda$ and $B$ into $\Lambda_{\rm UV}$ and $\alpha_{\rm UV}$ for a particular gauge group, and in Fig.~\ref{fig:1} we show the values of $\alpha_{\rm UV}$ if $\Lambda_{\rm UV}=10^{16}~\GeV$, for a supersymmetric ${\rm SU}\left(3\right)$ theory. For this model, the grand unified theory motivated value $\alpha_{\rm UV} \simeq 1/24$ requires an extremely small branching fraction. Values of $\Lambda$ around the DM self interaction bound, $\simeq 0.1~\GeV$, correspond to relatively small UV gauge couplings. Using Eq.~\eqref{eq:run}, the plotted values of $\alpha_{\rm UV}$ can be modified for other values of $\Lambda_{\rm UV}$ via
\beq \label{eq:convg}
\frac{1}{\alpha_{\rm UV}\left(\Lambda_{\rm UV}\right)} = \frac{1}{\alpha_{\rm UV}\left(10^{16}~\GeV\right)} - \frac{9}{2 \pi}\log\left(\frac{10^{16}~\GeV}{\Lambda_{\rm UV}} \right) ~,
\eeq
for an ${\rm SU}\left(3\right)$ sector, where $\alpha_{\rm UV}\left(10^{16}~\GeV\right)$ is the value shown in Fig~\ref{fig:1}. Lowering $\Lambda_{\rm UV}$ allows for glueball DM with slightly larger UV gauge couplings.

While we have assumed that the glueballs are stable this is not necessarily the case. Decay before big bang nucleosynthesis (BBN) is possible from a dimension 8 operator in the low energy effective theory
\beq
\mathcal{L}_{\rm IR} \supset \frac{\Lambda^3}{f_a^4} G \left(\partial_{\mu} a\right)\left(\partial^{\mu}a\right)~,
\eeq
coupling a glueball $G$ and an axion $a$, if $\Lambda$ is large and the axion decay constant $f_a$ is small \cite{Halverson:2016nfq}. For example, if $\Lambda \simeq 10^6~\GeV$, $f_a \lesssim 10^{10}~\GeV$ is needed, far below the typical values in string theory, with smaller decay constants required if the confinement scale is lower. If the glueballs do not decay before BBN but have a lifetime shorter than the age of the universe, their initial abundance must be small so that the universe is radiation dominated during BBN, and to evade 
observational constraints on late time entropy injection \cite{Scherrer:1987rr} (dark radiation limits can also be relevant since the axions produced are relativistic). Meanwhile for larger $f_a$ the glueballs are cosmologically stable and form a component of DM. If $\Lambda \simeq 10^6~\GeV$ this corresponds to $f_a \gtrsim 10^{12}~\GeV$, and for $f_a \sim 10^{16}~\GeV$ all glueballs with $\Lambda \lesssim 10^{10}~\GeV$ are stable \cite{Halverson:2016nfq}.

Glueball decays to the visible sector are also possible. Heavy states charged under both the visible and hidden sector gauge groups lead to a dimension 8 operator through which glueballs can decay to photons \cite{Boddy:2014yra}, and decays to the SM Higgs are possible through a dimension 6 Higgs portal \cite{Juknevich:2009gg}. These can allow the relic density constraint to be evaded, however typically a new intermediate mass scale must be introduced so that the decays happen sufficiently fast. In particular, for a Higgs portal operator suppressed by the string scale $\sim 10^{16}~\GeV$ decays only happen after BBN if $\Lambda\gtrsim 10^{11}~\GeV$ (the photon coupling leads to slower decays with the same suppression scale). Meanwhile unless $\Lambda\gtrsim 10^5 ~\GeV$ the glueballs have a lifetime shorter than the age of the universe so can not be DM. In this case their initial abundance must be significantly below that of the DM to avoid observational constraints on late time visible sector energy injection \cite{Jedamzik:2006xz}, and even smaller values of $B$ than if the glueballs are DM are needed. Consequently, decays do not appear to be a generic solution for a UV theory with many hidden gauge sectors. Glueball decays to hidden sectors may also be possible, but require the introduction of new light states, and the relic density of these will itself be challenging to accommodate.

\section{Glueball dark matter after matter domination} \label{sec:low}

We now turn to the main focus of our work, theories with a non-thermal cosmological history and a low reheating temperature after the final period of matter domination.\footnote{Conventional dark matter possibilities in such scenarios have been studied extensively, for example \cite{Giudice:2000ex,Kane:2015qea,Kane:2015jia}, and weakly coupled hidden sector gaugino dark matter in non-thermal cosmologies has been considered in \cite{Blinov:2014nla}.}
In contrast to scenarios with a high final reheating temperature, in these models the glueball sector is often far from chemical equilibrium, affecting the  relic abundance. As before we assume a hidden sector containing just gauge interactions, with confinement scale $\Lambda$.  

From the point of view of string theory, which generically predicts hidden sectors,
there does not appear to be a strong reason why a particular sector will be reheated
much more or less than any other. For this to occur, one would have to arrange for
the lightest modulus field to couple most strongly to the SM sector,
which seems non-generic since any such modulus will be a linear combination of mass
eigenstates all with roughly equal masses and lifetimes. Further, in UV theories the  branching fraction of moduli to vector superfields is often comparable or larger than that to any chiral superfields present  \cite{Nakamura:2006uc}. Even if absent at leading order, moduli couplings to pure gauge sectors are generated at one loop from the super-Weyl anomaly \cite{Kaplunovsky:1994fg,Moroi:1999zb,Endo:2007ih} (although these are Planck suppressed so could be somewhat smaller than any string scale suppressed couplings present). As a result the pure gauge nature of such hidden sectors does not automatically lead to it getting a small energy density after the modulus decay in typical models.

Suppose the last stage of reheating is through the decay of a gravitationally coupled string modulus $X$, with mass $m_X$. Once the temperature of the universe drops to $T_{\rm os} \simeq \left(m_{X} M_{\rm Pl}\right)^{1/2}$, where $M_{\rm Pl}$ is the reduced Planck mass, this will start oscillating and quickly dominate the energy density \cite{Dine:1995kz}. At much later times the modulus will decay, with a rate that is parametrically
\beq
\Gamma_X \simeq \frac{m_X^3}{M_{\rm Pl}^2} ~,
\eeq
and to our level of precision the finite time over which the modulus decay occurs can be neglected. The modulus decay happens when the Hubble parameter $H_0$ is approximately equal to $\Gamma_X$, and the total energy density is of order
\beq
\rho_{\rm tot} \simeq \frac{m_X^6}{M_{\rm Pl}^2} ~,
\eeq
therefore there is a long period of matter domination, during which any energy with the equation of state of radiation is diluted \cite{Kane:2015qea}.  In Section~\ref{sec:resid}, we study the possibility that energy in the glueball sector before matter domination could play a significant role after. There we show that over almost all of the parameter space of interest in this section the dilution is sufficient that the energy in glueball sector before matter domination is irrelevant afterward, and only the dynamics after the modulus decay needs to be considered.

To avoid an overabundance of glueballs, the visible sector must be reheated more strongly by the modulus decays than the hidden sector, so the visible sector reheat temperature $T_{\rm RH}$ is $\simeq \rho_{\rm tot}^{1/4}$. Constraints from BBN require $T_{\rm RH}$ is above an MeV, and as a result $m_X \gtrsim 10^4~\GeV$. Similarly to Section~\ref{sec:high} we define a parameter $B$ that measures how asymmetric the modulus decay is 
\beq
B= \frac{g_{v} \rho_g}{g_{g} \rho_v} \simeq \frac{g_g \rho_g}{g_{v} \rho_{\rm tot}} ~.
\eeq
Here $g_g$ is the hidden sector gluon effective number of degrees of freedom, $g_{v}$ the effective number of visible sector degrees of freedom that are light enough for the modus to decay to, and the energy densities are evaluated immediately after the modulus decay. If $m_X \gtrsim \Lambda$ decays to the hidden sector are not kinematically suppressed \cite{Allahverdi:2002nb}, and significant reheating is expected unless the modulus coupling to it is small relative to that to the visible sector.\footnote{Due to the modulus' very weak couplings, its decay is perturbative, and non-perturbative effects such as preheating are not relevant. In other models, with more strongly coupled scalars, non-perturbative dynamics could potentially have interesting effects.}

Immediately after the modulus decay, the gluon number density is 
\beq \label{eq:n0}
n_0 \simeq \frac{g_g}{g_v} B n_{X} \simeq \frac{g_g}{g_{v}} B \frac{m_X^5}{M_{\rm Pl}^2}~,
\eeq
where $n_X = \rho_{\rm tot}/m_X$ is the effective number density of individual $X$ quanta making up the coherently oscillating modulus field at the time of decay.  For simplicity, we assume that $g_v$ is also the visible sector effective number of degrees of freedom once this is thermalised to $T_{\rm RH}$ (deviations from this assumption are equivalent to an $\OO\left(1\right)$ rescaling of $B$). We also assume  that the visible sector thermalises immediately, which is sufficient for our present accuracy, though in reality this is a complex process \cite{Drewes:2014pfa}. 

\begin{figure}
	\begin{center}
		\includegraphics[width=.6\textwidth]{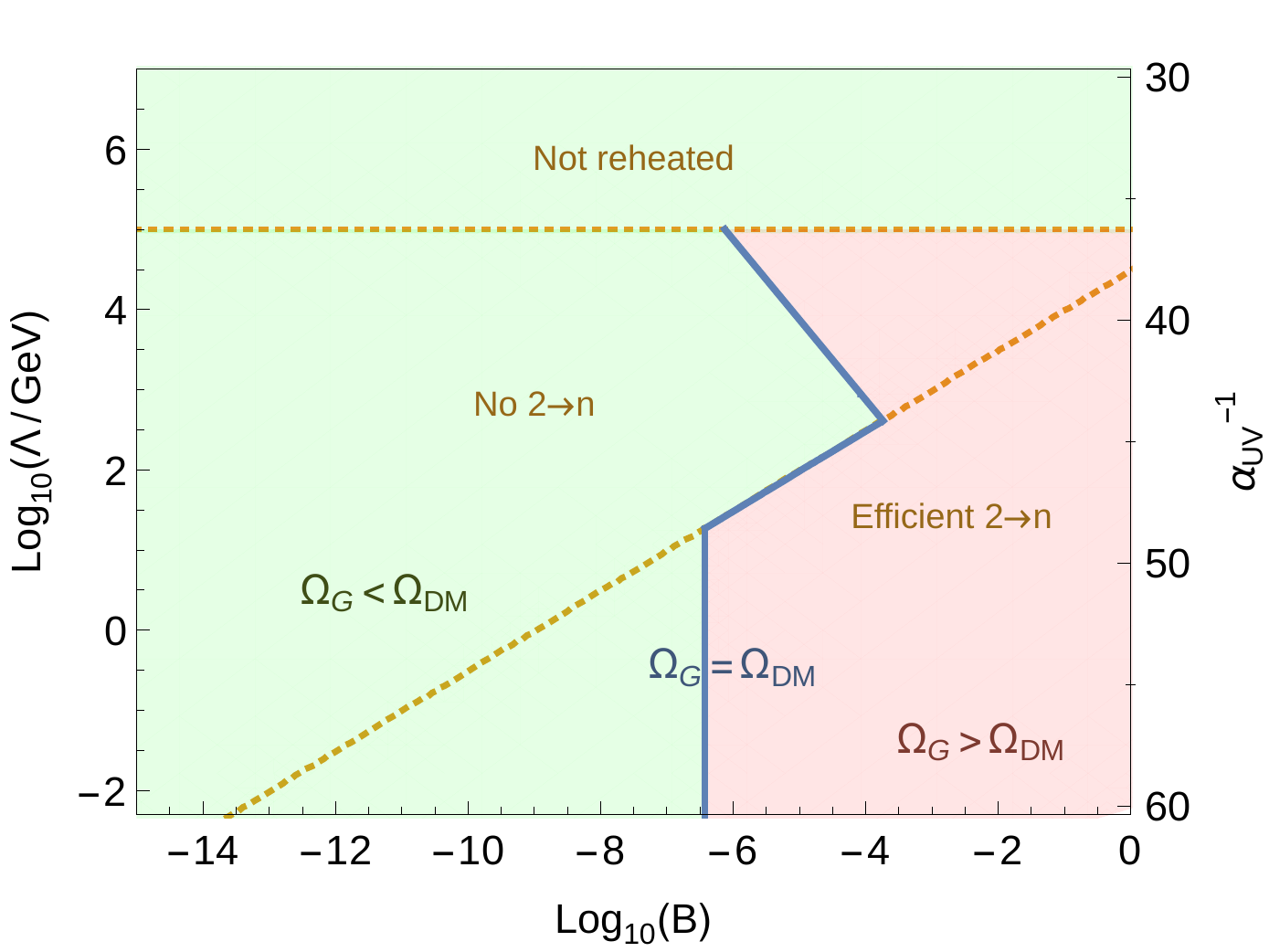}
	\end{center}
	\caption{
		The glueball DM parameter space for an ${\rm SU}\left(3\right)$ hidden sector with a period of late time matter domination and a low final reheat temperature, due to a gravitationally coupled modulus with mass $m_X = 10^5~\GeV$. The correct relic density is indicated by a blue line and requires a modulus branching ratio to the glueball sector $B$ significantly smaller than $1$, but less extreme than in theories with high scale reheating. Regions shaded green have an under abundance of glueball dark matter. For large $\Lambda$, $2\rightarrow n$ scatterings are never efficient, while for smaller values, the glueball number density increases until they become non-relativistic. If $\Lambda \gtrsim m_{X}$ the glueball sector is not reheated. Close to the boundaries between regions the dynamics are more complicated, beyond the scope of our approximations. The UV gauge couplings corresponding to $\Lambda$ are also shown, defined at a scale $10^{16}~\GeV$, and assuming the glueball sector is 
		supersymmetric above $10^6~\GeV$.
	}
	\label{fig:2}
\end{figure}

The initial gluon number density, Eq.~\eqref{eq:n0}, is typically much smaller than $\Lambda^3$, regardless of the temperature that a thermalised system with the same total energy would have. Consequently, they form glueballs in a time $\sim 1/\Lambda$, fast compared to other relevant processes, and similarly to Section~\ref{sec:high} we take that each gluon leads to one glueball. Depending on the details of the strong interactions this could plausibly change by a factor expected to be $\lesssim 10$.  Since $m_X$ is usually large compared to $\Lambda$ these are ultra-relativistic, and have very low number density compared to a thermalised system. The subsequent dynamics depends on the particular model, and the viable parameter space splits into different scenarios. The range of possibilities is summarised in Fig.~\ref{fig:2} for models 
with $m_X=10^5~\GeV$ and $m_X =10^6~\GeV$.

There is a boundary between dynamical regimes depending on whether the glueballs that are first produced have a high enough number density that $2 \rightarrow n$ interactions are fast compared to Hubble expansion, with $n >2$. We assume the $2 \rightarrow 3$ cross section is parametrically given by the geometrical expectation, with a suppression in the large $N$ limit for an ${\rm SU}\left(N\right)$ theory \cite{Forestell:2016qhc},
\beq \label{eq:23cs}
\left<\sigma v\right>_{2\rightarrow 3} \simeq \frac{\left(4 \pi\right)^3}{N^6} \frac{1}{\Lambda^2}~. 
\eeq
Taking this value despite the high center of mass energies of typical collisions is motivated by LHC data, which shows that the elastic and inelastic proton-proton cross sections are close to constant up to a high center of mass energy $\sqrt{s} \sim 7~\TeV$  \cite{Fagundes:2012rr}. 
The cross section for $2\rightarrow n$ processes with $n>3$ is parametrically similar to Eq.~\eqref{eq:23cs}, but more suppressed in the large $N$ limit. Additionally, once the glueballs have low kinetic energy $2\rightarrow 3$ interactions are dominant due to kinematics. Therefore it is a reasonable approximation to take the rate of production of extra glueballs by all $2\rightarrow n$ scatterings to be fixed by the the rate of $2\rightarrow 3$ interactions.

For $2 \rightarrow 3$ scattering to be faster than Hubble expansion requires $n_0 \left<\sigma v\right>_{2\rightarrow 3} \gtrsim 3 H_0$, where $n_0$ is given by Eq.~\eqref{eq:n0}, which is equivalent to
\beq \label{eq:initscat}
\begin{aligned}
B \frac{\left(4 \pi\right)^3}{N^6} \frac{g_g}{g_{v}} \frac{m_X^2}{\Lambda^2} && \gtrsim 3~.
\end{aligned}
\eeq
If Eq.~\eqref{eq:initscat} is not satisfied at early times then efficient scattering will not happen later either, since $n_g\sim 1/a^3$ while $H\sim 1/a^2$ where $a$ is the scale factor of the universe. If $2\rightarrow n$ scattering is not fast, the yield of glueballs will only increase from $n_0 /s_0$ by a factor less that $\mathcal{O}\left(1\right)$, where $s_0$ is the initial visible sector entropy.\footnote{Such a scenario is reminiscent of models in which a WIMP dark matter relic abundance is set directly by a modulus decay if subsequent interactions are slow compared to Hubble \cite{Gelmini:2006pw,Allahverdi:2012gk,Kane:2015jia}.} Meanwhile, the glueball kinetic energy will redshift away, resulting in a late time yield $y_{\infty}$ given by
\beq \label{eq:noscattery}
y_{\infty} \simeq \frac{n_0}{\frac{2 \pi^2}{45} g_v T_{\rm RH}^3} \simeq B \frac{g_g}{ g_v^{5/4}} \sqrt{\frac{m_{X}}{M_{\rm Pl}}} ~.
\eeq
This is suppressed relative to the high reheat temperature scenario by a potentially large factor, caused by the high proportion $m_X /\Lambda$ of the glueballs' energy remaining as kinetic energy (this is also a change because the yield is set immediately after reheating, rather than when the hidden sector temperature is $\sim \Lambda$).

The contribution of glueballs from an ${\rm SU}\left(N\right)$ gauge sector to the DM relic abundance in this scenario is
\beq\label{eq:noscatterb}
\frac{\left(\Omega h^2\right)_{G}}{\left(\Omega h^2\right)_{\rm DM}} \simeq \left(N^2-1\right) \left(\frac{B}{3\times 10^{-5}}\right)\left(\frac{m_X}{10^5~\GeV}\right)^{1/2} \left(\frac{\Lambda}{10^4~\GeV}\right)~,
\eeq
where $\Lambda$ has been normalised relative to a fairly high scale, since this is typically needed for $2\rightarrow n$ scattering to be inefficient (that is, Eq.~\eqref{eq:initscat} to not be satisfied). The glueball kinetic energy is initially high, so there must be sufficient time after their production for this to redshift away, otherwise they will violate observational constraints on hot dark matter. Careful analysis would require calculating the power spectrum, however we can estimate the constraints since the current free-streaming velocity is required  to be $\lesssim 3\times 10^{-8}$ by Lyman-$\alpha$ observations \cite{Viel:2005qj,Boyarsky:2008xj,Viel:2013apy} (dwarf spheroidal galaxies lead to similar constraints \cite{Boyarsky:2008ju,Horiuchi:2013noa}). Therefore, by the time the visible sector temperature reaches its present day value $T_t$, the glueball velocity $v_G$ (the idealised velocity left over from the Early Universe, not including the velocity which would subsequently be obtained through gravitational clustering) must satisfy
\beq
\begin{aligned}
v_G \simeq \frac{T_{t}}{T_{\rm RH}} \frac{m_X}{\Lambda} &\lesssim  3\times 10^{-8}  \\
\imp \left(\frac{\Lambda}{10^5~\GeV}\right) \left(\frac{m_X}{10^5~\GeV}\right)^{1/2} &\gtrsim 10^{-3.5} ~.
\end{aligned}
\eeq

This is safe over all of the interesting parameter space, since $\Lambda$ cannot be too much smaller than $m_X$ otherwise scattering will be efficient from Eq.~\eqref{eq:initscat}.

The alternative is that Eq.~\eqref{eq:initscat} is satisfied, and $2 \rightarrow n$ scattering is fast when the initial glueballs are produced by the modulus decay, in this case the number density of glueballs increases dramatically. Since the extra glueballs from scatterings themselves take part in further interactions causing a runaway process, the boundary between this regime and the previous one is sharp.

A thermalised sector with the same energy density as the glueballs will have temperature higher than $\Lambda$ if
\beq \label{eq:tempaboveL}
\left(\frac{10}{3 g_v}\right)^{1/4} B^{1/4} \frac{m_X^{3/2}}{M_{\rm Pl}^{1/2}} \gtrsim \Lambda~.
\eeq
 The dynamics of thermalisation are complex, and will be important in models close to this boundary. However, over most of parameter space, it is a reasonable approximation that if  Eq.~\eqref{eq:tempaboveL} is satisfied, and there is efficient $2 \rightarrow n$ scattering, a thermalised gluon plasma will form.\footnote{From  Eq.~\eqref{eq:initscat} for all the parameter space with $\Lambda \gtrsim 0.1~\GeV$ and $m_{X}\lesssim 10^{8.5}~\GeV$, if Eq.~\eqref{eq:tempaboveL} is satisfied, $2 \rightarrow n$ scattering will also be efficient.} 
If such a plasma forms, the ratio of the hidden and visible sector entropy density is 
\beq
\frac{s_g}{s_v} \simeq \frac{g_g}{g_{v}} B^{3/4}~,
\eeq
which is exactly the high reheating temperature scenario, and the glueball relic density is as calculated in Section~\ref{sec:high}. As the mass of the modulus is increased more of the parameter space is in this regime.

If the temperature of an equivalent thermalised system is below $\Lambda$, the glueball number density will increase fast until they become non-relativistic. The number density when this happens is approximately
\beq \label{eq:nnr}
n_{\rm nr} \simeq B \frac{g_g}{g_{v}} \frac{m_X^6}{2 M_{\rm Pl}^2 \Lambda} ~.
\eeq
Subsequently production of new glueballs slows down, as fewer collisions have enough center of mass energy to generate extra glueballs. Kinetic energy is distributed approximately evenly between glueballs, because $2\rightarrow 2$ scattering is more common than other processes once only a fraction of collisions can produce new glueballs. In particular once the glueballs have a velocity distribution close to Maxwell-Boltzmann, $2\rightarrow 3$ processes are suppressed by $\sim \sqrt{\Lambda/E_G} \exp\left(-\Lambda/E_G\right)$ where $E_G$ is the average glueball kinetic energy.

The glueball number density continues increasing from $n_{\rm nr}$ (by up to a factor of $2$) until $2\rightarrow 3$ scattering either becomes inefficient compared to Hubble expansion, or suppressed enough that it happens at the same rate as $3\rightarrow 2$ processes. In both cases $E_G$ is typically small $\lesssim\frac{1}{10} \Lambda$ when this happens (the former because the number density has increased dramatically compared to immediately after the modulus decay, the later because the number density is still significantly below $\Lambda^3$). As a result, almost all of the glueball kinetic energy is converted to glueball mass, and the number density is $\simeq 2 n_{\rm nr}$, defined in  Eq.~\eqref{eq:nnr}. 

If $3 \rightarrow 2$ processes become efficient these can reduce the glueball number density slightly, analogously to the dynamics in the models of Sector~\ref{sec:high}. In Appendix~\ref{sec:details} we show that the effect on the final number density is small, which remains well approximated by $\simeq 2 n_{\rm nr}$. This is because the glueball number density is much smaller than $\Lambda^3$, so begins tracking the equilibrium value once the glueball temperature is fairly low, and  $3 \rightarrow 2$ processes only reduce the yield for a short time. 
Using the approximation $n \simeq 2 n_{\rm nr}$, the late time yield is
\beq \label{eq:yapprox3}
y_{\infty} \simeq B \frac{g_{g}}{ g_v^{5/4}} \frac{m_{X}^{3/2}}{M_{\rm Pl}^{1/2} \Lambda}~,
\eeq
and unlike in other parts of parameter space, the relic density is independent of $\Lambda$,
\beq \label{eq:bnrapprox}
\frac{\left(\Omega h^2\right)_{G}}{\left(\Omega h^2\right)_{\rm DM}} \simeq  \left(N^2-1\right) \left(\frac{B}{3\times 10^{-6}}\right)  \left(\frac{m_X}{10^5~\GeV}\right)^{3/2} ~,
\eeq
for an ${\rm SU}\left(N\right)$ sector, assuming $g_{v}$ is the high temperature SM value.

Comparing Fig.~\ref{fig:2} with the thermal cosmology scenario plotted in Fig.~\ref{fig:1}, larger values of $B$  are viable in theories with late time matter domination, especially  when $\Lambda$ is high. The contour corresponding to the correct relic abundance in models with low reheating temperature meets that in the high temperature reheating case when the glueball sector has an equivalent thermalised temperature equal to $\Lambda$.\footnote{There is a small correction from approximating the final yield as $2 n_{\rm nr}$ in the matter dominated scenario.} This provides the link between the two scenarios, and as discussed around Eq.~\eqref{eq:tempaboveL} is exactly the point where the dynamics move into a different regime. In contrast, to our level of approximation, the boundary between $2\rightarrow n$ interactions being efficient or not is effectively discontinuous, because of the run-away nature of the scattering rate.  We also note that while the relative energy in the glueball sector can be larger in non-thermal cosmologies, the hidden sector phase transition happens during matter domination. The intensity of gravitational waves is therefore diluted, and these are again unobservable with planned detectors.

In Fig.~\ref{fig:2} we also show the UV gauge couplings corresponding to $\Lambda$ for a supersymmetric ${\rm SU}\left(3\right)$ theory, using Eq.~\eqref{eq:run}. As in Fig.~\ref{fig:1} the high scale is taken as $\Lambda_{\rm UV}=10^{16}~\GeV$, with the couplings at other UV scales related by Eq.~\eqref{eq:convg}.  From a UV perspective, the glueball relic abundance is more favourable than in the high reheat temperature case. For relatively large gauge couplings $\alpha_{\rm UV} \simeq 1/24$, which would vastly overclose the universe with a high reheat temperature, glueball production is simply blocked by $\Lambda \gtrsim m_{X}$ for typical moduli masses. Meanwhile, for slightly smaller but still plausible gauge couplings, glueball DM is possible. This requires a fairly small modulus branching fraction, but values closer to $\OO\left(1\right)$ than in the high reheat temperature case are allowed.

Glueball DM can lead to a variety of experimental signals, although these are model dependent and not guaranteed to be present. Since the glueballs must interact very weakly with the visible sector so that they are sufficiently stable, direct detection signals are unobservable in viable models. However, higher dimensional operators could lead to indirect detection signals, such as through decays to monochromatic photon lines \cite{Soni:2016gzf}. 
The strongly interacting nature of the interactions could also allow for the formation of dark boson stars \cite{Soni:2016gzf}, which may be observable through microlensing \cite{Dabrowski:1998ac}. These signatures are not unique to glueball DM, however if combined with evidence for self-interactions of the expected strength given the inferred DM mass, glueballs would be a compelling candidate.

A period of late time matter domination can have observable effects, for example on BBN and the cosmic microwave background (a review may be found in e.g. \cite{Kane:2015jia}). If evidence for a matter dominated period was found and DM properties (such as self interactions) compatible with glueballs measured  this would hint towards the particular scenario we study. Unfortunately such signals are only observable in parts of parameter and model space. Additionally, during matter domination the growth of density perturbations is faster than during radiation domination, potentially modifying the distribution of DM on small scales. For non-thermal WIMP models there is no significant impact on the final distribution since the DM is kinetically coupled to the radiation bath after reheating \cite{Fan:2014zua}, however this is not the case in glueball models. Consequently it is possible this could lead to observable effects that are hard to reproduce in other theories, and we hope to investigate this in future work. More generally, although not evidence for glueball DM, experimental signs of hidden sectors for example from dark radiation or additional dark photons, would increase the plausibility of the models we consider. While the lack of a unique clean experimental signature is unfortunate, it does not decrease the theoretical reasonableness of the scenario of interest.

\section{Hidden sector gluinos} \label{sec:gluino}

Models containing hidden sector gluinos $\tilde{g}$ with mass $m_{\tilde{g}}$ below that of the moduli are also well motivated from string theory. If the lightest modulus has a significant branching fraction to these, they can be cosmologically relevant (their possible role in models with a high reheating temperature has previously been studied \cite{Feng:2011ik,Boddy:2014yra,Boddy:2014qxa}). Alternatively it is plausible that supersymmetry is broken strongly in the glueball sector resulting in kinematically inaccessible heavy gluinos. 

As in the pure glueball case, immediately after the modulus decays the number density of gluons and gluinos is far below the chemical equilibrium value, and in the parameter space of interest the hidden sector energy density corresponds to a temperature below $\Lambda$ regardless. Therefore the gluinos quickly form colour neutral bound states, similarly to the gluons forming  glueballs, either $\tilde{g} \tilde{g} $ or a glueballino $\tilde{G}= \tilde{g} g$.\footnote{The dynamics of these are reminiscent of those of previously considered visible sector glueballinos \cite{Keung:1983wz,Nussinov:1996qf,Raby:1998xr,Farrar:1994xm,Kauth:2009ud} and gluino gluino bound states  
\cite{Kuhn:1983sc,Goldman:1984mj,Kauth:2009ud}.} The relative abundances of these depends on the 
details of strong coupling physics. However, it is reasonable to assume that the initial number densities of the two are similar, or that glueballinos are dominantly produced if the gluinos are relatively heavy $m_{\tilde{g}} \gg \Lambda$. 

The gluino constituents of $\tilde{g}\tilde{g}$ states annihilate quickly to gluons, therefore this fraction of the system's energy is equivalent to if it was transfered directly to gluons by the modulus decay. However, glueballinos are stable because they are the lightest fermionic state in the sector. We define a modulus branching fraction to gluballinos
\beq
B_{\tilde{G}}=  \frac{r g_{\tilde{g}} \rho_g}{g_{v} \rho_{\rm tot}} \equiv   \frac{g_{\tilde{G}} \rho_g}{g_{v} \rho_{\rm tot}} ~,
\eeq
where $r \sim 1$ is the average number of glueballinos produced per gluino, and $g_{\tilde{G}}$ is an effective parameter. The gluballino mass is typically $m_{\tilde{G}} \sim m_{\tilde{g}}$ if this is larger than the confinement scale. Otherwise, $m_{\tilde{G}}\sim \Lambda$, and the dynamics are very similar to simply having a second glueball in the theory, and we do not consider such cases further. 

As long as glueballino scatterings are rare compared to a timescale $1/\Lambda$ they remain as composite bound states. Precise calculations are not possible in this regime, but we can make estimates for the cross sections of various processes similarly to \cite{Mohapatra:1997sc} and \cite{Kang:2006yd} (related discussion of visible sector gluino cosmology can be found in \cite{ArkaniHamed:2004fb}). 

A large fraction of the glueballino mass is concentrated in a small region of size $\sim 1/m_{\tilde{g}}$, and is surrounded by a much larger gluon cloud with size $\sim 1/\Lambda$. If the center of mass kinetic energy of a collision is below $\Lambda$, annihilation of two glueballinos can proceed through the formation of a bound state. Production of a bound state has cross section $\sim 1/\Lambda^2$ \cite{Kang:2006yd}, and leads to a system with binding energy $\sim \Lambda$. The subsequent dynamics are uncertain, but it is likely that the constituent gluinos typically annihilate before the bound state dissociates.  We therefore approximate the cross section for glueballino annihilation from low energy collisions as 
\beq
\left<\sigma v\right>_{\tilde{G} \tilde{G} \rightarrow g g} \simeq \frac{ v_{\tilde{G}}}{\Lambda^2} ~,
\eeq
where $v_{\tilde{G}}$ is the glueballino velocity. There is likely to be extra suppression in the large $N$ limit of an ${\rm SU}\left(N\right)$ theory, but this is beyond our present analysis.

In contrast, when the center of mass energy kinetic energy of a glueballino-glueballino collision is above $\Lambda$ (but still in a scenario where the number density of the sector is sufficiently low that they are composite states not gluinos), bound state formation is suppressed. The annihilation cross section is instead expected to vary parametrically as
\beq
\left<\sigma v\right>_{\tilde{G} \tilde{G} \rightarrow g g}   \sim \frac{\alpha\left(\sqrt{s}\right) v_{\tilde{G}} }{m_X^2}~,
\eeq
where $\alpha\left(\sqrt{s}\right)$ is the hidden sector gauge coupling at the energy scale $\sqrt{s}$, motivated by the requirement that the gluino cores of the gluballinos must overlap for annihilation. Meanwhile, the cross section for elastic scattering of glueballinos with each other or with glueballs, or inelastic scattering producing extra glueballs, depends on the energy transfered in the process. Events with small momentum transfer $\sim \Lambda$, have an unsuppressed cross-section $\sim 1/\Lambda^2$, and processes involving a larger momentum transfer are expected to have a more suppressed cross section.

With these approximations, the cosmological effects of gluinos can be described qualitatively, and the parts of parameter space for which they are important estimated. For simplicity we assume the modulus branching fraction to gluinos is equal to that to gluons. This is well motivated in UV models in which the modulus couples to the hidden sector through the gauge kinetic function  \cite{Nakamura:2006uc}. The effect of altering the relative branching fraction can be straightforwardly traced through our analysis.

In models in which glueballs never have efficient scattering, the same is also true for glueballinos. From Eq.~\eqref{eq:initscat}, this scenario typically occurs when $\Lambda$ is large and $B$ small. The rate of glueballino annihilation is suppressed initially since they are highly relativistic, and even interactions with small center of mass energy have a cross section parametrically the same as for glueball glueball scattering. Consequently, the glueballino yield remains fixed at the value immediately after modulus decay, similarly to Eq.~\eqref{eq:noscattery}, and glueballinos make up a fraction of the observed DM relic abundance
\beq \label{eq:gluinob}
\frac{\left(\Omega h^2\right)_{\tilde{G}}}{\left(\Omega h^2\right)_{\rm DM}} \simeq g_{\tilde{G}} \left(\frac{B_{\tilde{G}}}{6\times 10^{-5}}\right)\left(\frac{m_X}{10^5~\GeV}\right)^{1/2} \left(\frac{m_{\tilde{G}}}{10^4~\GeV}\right) ~.
\eeq
Due to their larger mass, glueballinos typically make up a greater fraction of the DM than glueballs in such models, leading to stronger constraints on the modulus branching fractions.

In theories with initial number densities high enough for glueball glueball $2 \rightarrow n$ scatterings to be efficient, glueballino scattering will be as well. This is because, as well as soft glueballino scattering having cross section $\sim 1/\Lambda^2$, the number density of glueballs grows rapidly, enhancing the glueballino glueball scattering rate. Elastic scatterings with glueballs decrease the average glueballino energy, since the glueballs themselves are losing energy fast through $2\rightarrow n$ processes. Until the glueballinos have lost almost all of their energy, annihilations are irrelevant compared to elastic collisions or events producing new low momentum glueballs, due to the momentum dependence of the cross sections.  

As a result, almost all of the glueballino initial kinetic energy is transfered to glueball states, increasing the effective value of $B$ for glueballs by an order 1 factor. We assume glueball glueball collisions  do not lead to more than an order 1 increase in the glueballino yield, which is reasonable if the the gluino mass is significantly larger than $\Lambda$ and is not 
too much smaller than $m_X$, so typical glueball collisions only have center of mass energy above $m_{\tilde{G}}$ for a relatively short fraction of the time for which $2\rightarrow n$ processes are active. The glueballino relic density is not very sensitive to this approximation, since in large parts of parameter space it is fixed by subsequent annihilations.

Once the glueballinos have kinetic energy $\lesssim \Lambda$, by which point they are highly non-relativistic with $v_{\tilde{G}} \sim \sqrt{\Lambda/m_{\tilde{G}}}$, annihilations become efficient compared to soft collisions. Provided the hidden sector energy density after modulus decay is $\lesssim \Lambda^4$, this regime is reached fast due to the increase in glueball number density, and at this time the glueballino number density is still approximately
\beq
n_{\tilde{G}}  \simeq \frac{g_{\tilde{G}}}{g_{v}} B \frac{m_X^5}{M_{\rm Pl}^2}~.
\eeq
Subsequently the glueballino number density can be reduced by annihilations if their number density is sufficiently high. Annihilations convert the glueballino energy to glueballs, leading to an $\OO\left(1\right)$ increase in yield of these. Otherwise, since very few collisions have enough energy to produce new gluinos, the yield will be constant and the contribution to the relic density given by Eq.~\eqref{eq:gluinob}.

The condition for annihilations to be important is $n_{\tilde{G}} \left<\sigma v\right>_{\tilde{G} \tilde{G} \rightarrow g g}  \gtrsim 3 H$, that is
\beq
\frac{g_{\tilde{G}}}{3 g_v} B_{\tilde{G}} \frac{m_{X}^3}{\Lambda^{3/2} m_{\tilde{G}}^{1/2}} \gtrsim 1 ~,
\eeq
which is satisfied over large parts, but not all, of the relevant parameter space.\footnote{This is similar to some non-thermal models of WIMP dark matter \cite{McDonald:1989jd,Giudice:2000ex,Catena:2004ba,Cheung:2010gj}.} If annihilations do take place,  they reduce the glueballino number density until it is \cite{Acharya:2009zt}
\beq \label{eq:gluinorelic}
n_{\tilde{G}} \simeq \frac{3H}{\left<\sigma v\right>_{\tilde{G} \tilde{G} \rightarrow g g} } \simeq \frac{3 m_X^3 \Lambda^{3/2} m_{\tilde{G}}^{1/2}}{M_{\rm Pl}^2 } ~.
\eeq
In this scenario the glueballino contribution to the relic abundance is
\beq \label{eq:gluinoome}
\frac{\left(\Omega h^2\right)_{\tilde{G}}}{\left(\Omega h^2\right)_{\rm DM}} \simeq 10^{-8} \left(\frac{\Lambda}{\GeV}\right)^{3/2} \left(\frac{m_{\tilde{G}}}{\GeV}\right)^{3/2} \left(\frac{10^5~\GeV}{m_X} \right)^{3/2} ~,
\eeq
while gluons produced by the annihilations will increase the glueball yield by an $\OO\left(1\right)$ factor. 
Parts of parameter space where initial scatterings are efficient typically have fairly low $\Lambda$, and a small glueballino relic density compared to glueballs. This is a result of the efficient glueballino annihilations, and is in contrast to if initial $2\rightarrow n$ glueball scattering is not fast.

Another possibility is that the glueball sector energy density is equivalent to a reheat temperature above $\Lambda$, in which case the gluons usually thermalise. If $T_g$ is below $m_{\tilde{G}}$ the glueballino freeze out is similar to before, but only begins once the hidden sector temperature drops to $\sim \Lambda$ and annihilations become efficient. As a result, the visible sector temperature will also have decreased, so the Hubble parameter is a factor $\frac{\Lambda^2}{T_g^2}$ smaller than when the modulus decays. The glueballino yield is therefore increased from Eq.~\eqref{eq:gluinoome} by a factor  $\frac{T_g}{\Lambda}$ (since the glueballino velocity has decreased). Meanwhile, if $T_{g}$ is above $m_{\tilde{G}}$ the gluinos typically reach chemical equilibrium and form a thermalised plasma rather than interact as glueballinos. Gluino freeze out takes place when the hidden sector temperature drops below $m_{\tilde{g}}$, as calculated in \cite{Boddy:2014yra}.\footnote{A second period of 
annihilations of the relic glueballinos could 
also happen in 
some models.} 
Models with such a high reheating temperature need a very small branching fraction to glueballs, and have dynamics equivalent to models previously studied in the literature with a high reheating temperature, and we do not consider them further.

Glueballinos have self-interactions that are potentially astrophysically relevant, and with interesting differences to the self-interactions of glueballs. For example they can interact through exchange of relatively light glueballs, and the details have been studied carefully in \cite{Boddy:2014yra}. However, for the theories we study, the DM is dominantly made up of glueballinos when $\Lambda$ is large, and in this regime self-interactions are not important at late times. This is not necessarily the case if the  modulus branching fraction to gluons and gluinos is very different, and it might be worthwhile to study the phenomenology of such scenarios. In particular, it might be that the dynamics of the interactions is sufficiently altered compared to commonly considered self-interacting dark matter models that there are potentially observable distinguishing astrophysical effects.

\section{Glueballs from before matter domination} \label{sec:resid}

In some parts of parameter space it is possible that energy in the glueball sector before matter domination could be cosmologically relevant afterwards, despite the dilution. The modulus begins oscillating at a time $t_{\rm os}$ when the hottest sector in the theory, which we assume to be the visible sector, has temperature $T_{\rm os} \simeq g_v^{-1/4} \left(m_X M_{\rm Pl}\right)^{1/2}$. At this point the glueball sector could be at a lower temperature, which we define as $B_i^{1/4} T_{\rm os}$, where  $B_i$ could be, for example, the inflaton branching fraction to the glueball sector, and does not have to coincide with the modulus branching fraction $B$.

For phenomenologically viable modulus masses, $T_{\rm os}$ is high and extremely small values of $B_i$ are not expected from UV physics, so we focus on scenarios in which the initial glueball sector temperature is above $\Lambda $. We also assume that the maximum temperature that the glueball sector would have if it began with zero temperature and was heated only by modulus decays is $\lesssim \Lambda$. The maximum temperature is parametrically $B^{1/4} m_{X}$, which is larger than the reheating temperature \cite{Giudice:2000ex}, and this assumption is valid over all the relevant parameter space. As a result, before the phase transition the glueball sector temperature is determined by its energy density at the start of matter domination. 

During matter domination  the temperature of the hidden sector gluons decreases as $1/a$, until reaching $\Lambda$ at a time $t_{\Lambda}$. The scale factor of the universe when this happens, relative to that at the beginning of matter domination, is
\beq \label{eq:atL}
\frac{a\left(t_{\Lambda}\right)}{a\left(t_{\rm os}\right)} \simeq \frac{B_i^{1/4} T_{\rm os}}{\Lambda} ~.
\eeq
At $t_{\Lambda}$ glueballs form with number density $\Lambda^3$, similarly to Section~\ref{sec:high}. To our current level of precision, it is reasonable to ignore the  effects of $3\rightarrow 2$ processes when the glueball number density is set, analogous to the approximation Eq.~\eqref{eq:yieldnrh}. 

\begin{figure}
	\begin{center}
		\includegraphics[width=.6\textwidth]{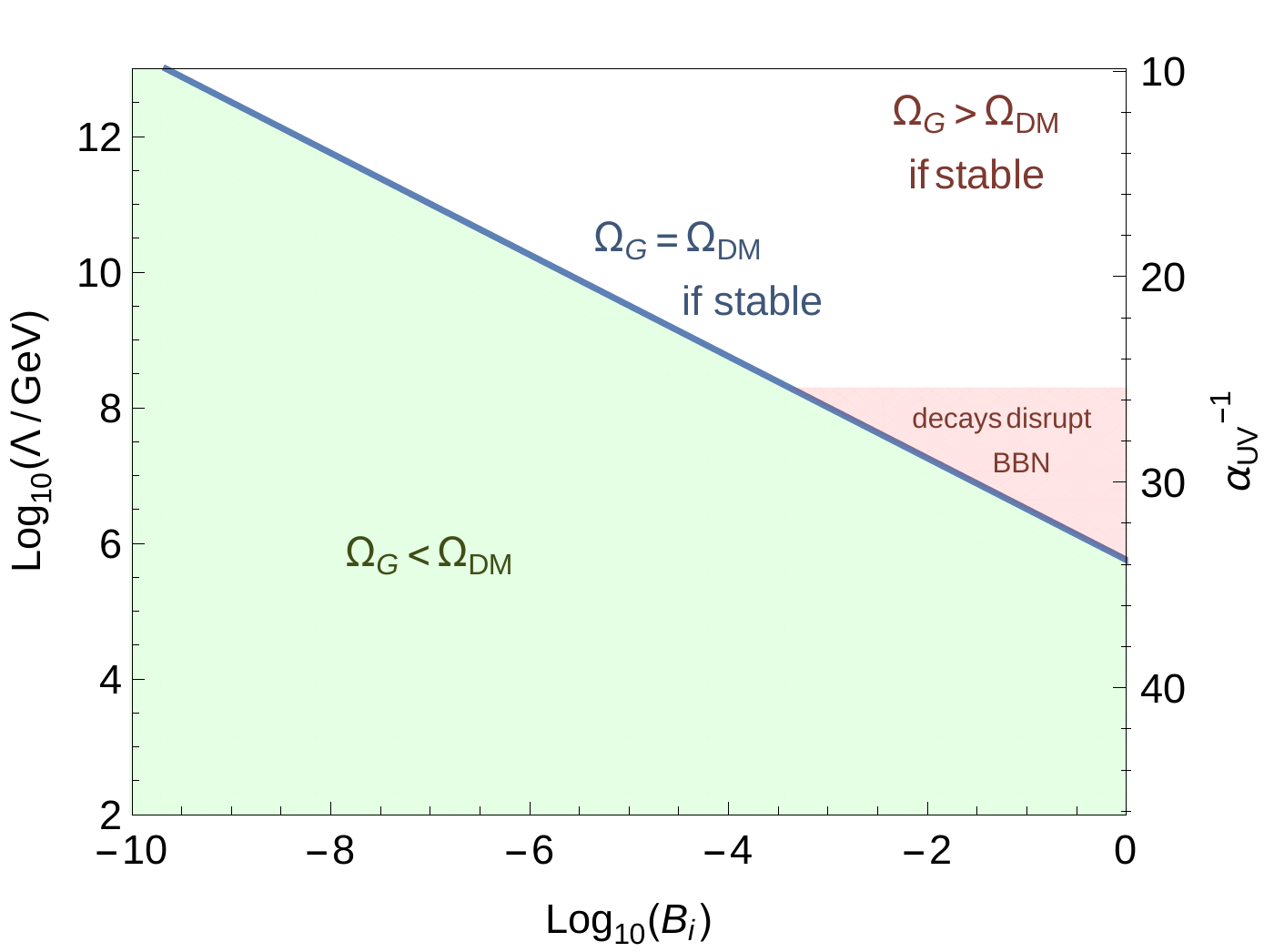}
	\end{center}
	\caption{The final relic abundance of glueballs produced from a phase transition that happens {\it during} a period of matter domination by a gravitationally coupled modulus with mass $m_X=10^5~\GeV$, as a result of energy in the glueball sector remaining from early times. The hidden sector is taken to be an ${\rm SU\left(3\right)}$ gauge group with confinement scale $\Lambda$, and $B_i = \left(T_{g}/T_v\right)^4$ is fixed by its temperature relative to the visible sector before matter domination. The corresponding gauge coupling at a scale $10^{16}~\GeV$ is shown assuming supersymmetry.	
If the glueballs are stable, their relic density matches the observed DM value on the blue contours. 		
		In parts of parameter space with $\Lambda \lesssim m_X$ extra glueballs can be produced by the modulus decay depending on its branching fraction as studied in Section~\ref{sec:low}.  For $\Lambda \gtrsim m_X$, fast glueball decays to the modulus are possible, depending on the UV theory. These prevent the glueballs being the DM, but avoid cosmological constraints, provided they occur before BBN (the region in which decays through dimension 6 operator generated at a scale $10^{16}~\GeV$ are dangerous is shaded red). 			
	}
	\label{fig:3}
\end{figure}

Between beginning oscillating and decaying the energy density in the modulus drops from $\left(m_X M_{\rm Pl}\right)^{2}$ to $m_X^6 / M_{\rm Pl}^2$.\footnote{We neglect the finite time over which the decays occur, which leads to an $\OO\left(1\right)$ correction to the results.} Therefore the scale factor when the modulus decays, at a time $t_d$, is
\beq
\frac{a\left(t_{d}\right)}{a\left(t_{\rm os}\right)} \simeq  \left(\frac{M_{\rm Pl}}{m_X}\right)^{4/3} ~,
\eeq
and the glueball number is diluted by a factor $\left(a\left(t_{\Lambda}\right)/ a\left(t_{d}\right) \right)^{3}$ to 
\beq
n_{g}\left(t_{d}\right) \simeq \frac{\xi\left(3\right)}{\pi^2} g_g \Lambda^3 \left( \frac{a\left(t_d\right)}{a\left(t_{\rm os}\right)} \frac{a\left(t_{\rm os}\right)}{a\left(t_{\Lambda}\right)} \right)^{-3} \simeq 0.1 \frac{g_g}{g_v^{3/4}} \frac{B_i^{3/4} m_{X}^{11/2}}{ M_{pl}^{5/2}} ~.
\eeq
After the modulus decays, reheating the visible sector to a temperature $T_{\rm d} \sim m_X^{3/2}/M_{\rm Pl}^{1/2}$, the glueball yield is approximately
\beq \label{eq:heavyyield}
y_g \simeq 0.1 \frac{g_g}{g_v} B_i^{3/4} \frac{m_X}{M_{\rm Pl}} ~,
\eeq
corresponding to a relic abundance
\beq \label{eq:relic}
\frac{\left(\Omega h^2\right)_{G}}{\left(\Omega h^2\right)_{\rm DM}} \simeq 0.02 \left(N^2-1\right) B_i^{3/4} \left(\frac{m_X}{10^5~\GeV}\right) \left(\frac{\Lambda}{10^5~\GeV}\right) ~.
\eeq
This is equivalent to noting that the modulus decay leads to an entropy injection that dilutes the glueball yield from the thermal value Eq.~\eqref{eq:yieldnrh} by a factor $T_{\rm os}/T_{\rm d} \sim m_X/ M_{\rm Pl}$  \cite{Scherrer:1984fd,Lazarides:1987zf,Lyth:1993zw}.

Over most of the parameter space studied in Section~\ref{sec:low}, the contribution Eq.~\eqref{eq:relic} is relatively small. In particular, if $B_i$ is taken equal to the modulus branching fraction to the glueball sector, the viable parameter space in Fig.~\ref{fig:2} is not constrained, apart from a small region with  $\Lambda > m_{X}$ and $B$ large. If $B_i =1$ independent of the value of the modulus branching fraction, the early relic abundance can be significant in models with $\Lambda$ close to typical values of $m_{X}$.

Additionally, the glueball relic density remaining after matter domination is sensitive to the details of the cosmological history.  While we have simply assumed one period of matter domination from a single modulus, string theory models are likely to include many moduli with a range of masses. During the evolution of the universe, the energy density will be dominated by a series of increasingly light moduli as heavier ones decay, which can result in energy density in glueballs from before matter domination being irrelevant even in parts of parameter space with large $\Lambda$. 
For example, a second heavier modulus with mass $m_H$ causes matter domination to begin earlier, and so a larger dilution from entropy injection. This suppresses the final glueball relic abundance by a factor $\sim \sqrt{m_H/m_X}$ compared to Eq.~\eqref{eq:relic}, assuming the lighter modulus starts oscillating during the first stage of matter domination.

If $\Lambda \gtrsim 2 m_X$ fast glueball decays to the modulus are also possible in many theories. In particular a dimension 6 operator in the in the UV Lagrangian
\beq \label{eq:decaytomod}
\mathcal{L_{\rm UV}} \supset \frac{X X}{M_{\rm UV}^2} {\rm Tr}\left(G_{\mu \nu} G^{\mu \nu} \right) ~,
\eeq
is allowed, where $M_{\rm UV}$ is a UV scale, for example the string scale, and $G_{\mu \nu}$ is the field strength of the hidden sector gauge group \cite{Halverson:2016nfq}. At energies below $\Lambda$ this leads to a coupling that is parametrically
\beq \label{eq:gdop}
\mathcal{L_{\rm IR}} \supset \frac{\Lambda^3}{M_{\rm UV}} G X X~,
\eeq
where $G$ is the glueball. For $M_{\rm UV}\sim M_{\rm Pl}$ the decay rate corresponding to such an operator is fast, and any glueballs produced cannot be the DM.\footnote{A dimension 5 operator $\frac{1}{M_{\rm UV}} G \partial X \partial X$ could also be written in the low energy effective theory, but does not come out of the underlying theory in a simple way \cite{Halverson:2016nfq}.} To have glueballs that are stable on timescales on the age of the universe, higher dimension operators of the form
\beq \label{eq:genmodd}
\mathcal{L_{\rm IR}} \supset \left(\frac{\Lambda}{M_{\rm UV}}\right)^{n} \Lambda G X X ~,
\eeq
with $n>2$, must  also typically be suppressed. For example, cosmologically stable glueballs with $M_{\rm UV} = M_{\rm Pl}$, $m_X=10^5~\GeV$, and $\Lambda=10^{9}~\GeV$ require the coefficient of the dimension 7 operator, Eq.~\eqref{eq:genmodd} with $n=3$, to be small. Particular UV completions might realise an exponential suppression of these operators, although we do not consider explicit scenarios.

Glueball decays to moduli can be phenomenologically beneficial in theories containing multiple hidden sectors, since they prevent glueballs with high confinement scales  over closing the universe. For $\Lambda \gtrsim 10^8~\GeV$ decays through the operator Eq.~\eqref{eq:gdop} happen before BBN for $M_{\rm UV}=10^{16}~\GeV$, and are cosmologically safe. While a dangerous region of parameter space remains for $B_i\sim 1$, sectors with large values of $\Lambda$ are significantly less constrained than in theories with no light moduli. Decays to axions are also possible, however these are expected to happen only through a dimension 8 operator, and for large axion decay constants $\sim 10^{16}~\GeV$ are irrelevant \cite{Halverson:2016nfq}.

Despite these caveats, we note that for large values of $\Lambda$ the glueball abundance produced during matter domination can match the observed relic abundance. In most such models $\Lambda \gg m_X$, so no glueballs are produced from the subsequent decay of the modulus, and Eq.~\eqref{eq:heavyyield} is the only contribution to the yield. In Fig.~\ref{fig:3} the parameter space that leads to the correct relic abundance is plotted, assuming a single modulus with mass $10^5~\GeV$ or $10^6~\GeV$. For motivated values of the UV gauge coupling, not far from the visible sector unification value, the correct relic abundance can be obtained for $B_i \simeq 10^{-6}$.

\section{Discussion} \label{sec:discuss}

Glueball dark matter is  well motivated from typical string compactifications, and can lead to interesting phenomenology \cite{Halverson:2016nfq}. 
For example, pure gauge hidden sectors seem to be common in heterotic \cite{Faraggi:1997dc,Lebedev:2006kn,Anderson:2013xka}, IIB \cite{Gmeiner:2005vz,Blumenhagen:2008zz}, M-theory \cite{Acharya:1998pm,Halverson:2015vta,Acharya:2016fge}, and F-theory \cite{Grassi:2014zxa,Halverson:2015jua,Taylor:2015xtz,Halverson:2016vwx} models. Meanwhile glueball self-interactions can be astrophysically relevant, and observational constraints require the hidden sector confinement scale is above approximately $100~\MeV$. 

In this paper we have shown that a non-thermal cosmology allows for viable theories with a more democratic final period of reheating compared to scenarios with a standard thermal cosmology. In models with a thermal cosmology and high reheat temperature a large initial entropy ratio between the visible and hidden sectors is needed to obtain the required glueball DM relic abundance (although this could be accommodated in some string constructions \cite{Harling:2008px}). The situation is worse for larger confinement scales, and if the hidden sector gauge couplings are similar to those of the visible sector at a scale $10^{16}~\GeV$, the difference in initial temperatures must be enormous. If the hope of glueball DM is given up, and the aim is simply to accommodate such gauge sectors, decays to the visible sector are possible (for larger couplings there could also be interesting collider signatures 
\cite{Juknevich:2009ji,Englert:2016knz}). These need new states at an intermediate mass scale, and potentially lead to constraints from energy injection to the visible sector. As a result, obtaining viable phenomenology from UV models that predict a high 
reheating temperature and multiple decoupled pure gauge sectors is potentially challenging.

In contrast, the inclusion of light gravitationally coupled moduli alleviates these problems, since the lightest modulus typically dominate the universe at late times, leading to a low final reheating temperature and changing the dynamics of hidden sector glueballs. The observed DM relic density still requires that the lightest modulus has a relatively small branching fraction to the hidden sector, but this is less extreme than in the case of a high reheat temperature. 
The glueball relic abundance is also less dependent on the hidden sector confinement scale, or equivalently the UV value of the gauge coupling, which is beneficial if an underlying theory is expected to contain multiple pure gauge hidden sectors with varying properties.

Additionally, hidden sectors with confinement scales above the mass of the lightest modulus are not reheated by the modulus decay. We have shown that they could still have a significant relic abundance from glueballs produced during the last period of matter domination, however this is model dependent, and the final abundance can be small in theories with multiple light moduli. Further, there is a dimension 6 operator that allows glueball decays to the light moduli, and this is usually fast enough to avoid cosmological constraints for confinement scales in the range motivated by unification of the visible sector gauge couplings. This is again in contrast to the high scale reheating temperature scenario, without lighter moduli, for which high confinement scales are potentially problematic. Unfortunately the scenario we study has no guaranteed experimental signature uniquely distinguishing it from other possibilities. However, depending on the details of an individual model, observational hints of both the glueball nature of the DM and a late time period of matter domination are possible.

Finally our discussion can also be interpreted in the opposite direction. A strong theoretical argument, for example from string theory, that typical UV completions of the visible sector should also include many disconnected pure gauge sectors might be possible in future. If combined with evidence that the branching ratio to all of these during the final period of reheating should not be extremely small, observations of the dark matter relic abundance would favour a period of late time matter domination. A natural candidate to generate this is the lightest modulus in the theory, which has a mass typically set by the gravitino mass. Consequently it is even tempting to interpret this as a hint that the gravitino mass (and therefore the scale of supersymmetry breaking) may not be so far above a TeV if the visible sector is UV completed by such a theory.

\subsection*{Acknowledgements}

MF  is  funded  by  the  European Research  Council  under  the  European  Union’s  Horizon 2020 program (ERC Grant Agreement no.648680 DARKHORIZONS).  BA  and  MF  are  supported  by  the  UK STFC Grant ST/L000326/1. This work was supported by a grant from the Simons Foundation (\#488569, Bobby Acharya).

\appendix

\section{$3\rightarrow 2$ freeze out in low reheat models} \label{sec:details}

In this appendix we discuss effects that can make an $\OO\left(1\right)$ difference to the glueball relic density in the non-thermal models studied in Section~\ref{sec:low}. In particular, these can modify the yield from the approximation Eq.~\eqref{eq:yapprox3} in the case in which $2\rightarrow 3$ processes increase the glueball number density immediately after the modulus decays. 

In this scenario the parameter space is split depending on whether the glueball number density becomes high enough for $3 \rightarrow 2$ processes to be efficient compared to Hubble expansion. When $2\rightarrow 3$ interactions become slow the glueball number density is close to $2 n_{\rm nr}$, and
to a good approximation $3 \rightarrow 2$ processes are active if
\beq
\begin{aligned} \label{eq:hubbor32}
4 n_{\rm nr}^2 \left<\sigma v\right>_{3\rightarrow 2} & < 3 H\left(T_{v0} \right) \\
\imp 10^{3} B^2 \frac{g_g^2}{g_{v}^2 N^6} \frac{m_X^9}{\Lambda^7 M_{\rm Pl}^2} & \lesssim 1~,
\end{aligned}
\eeq
using the parametric form of the cross section in Eq.~\eqref{eq:32scal}. If this is not satisfied, the final abundance is just fixed by the value of $2 n_{\rm nr}$.

Meanwhile if Eq.~\eqref{eq:hubbor32} is satisfied then $3\rightarrow 2$ processes become efficient, and the system will typically reach full chemical equilibrium. When the glueballs have kinetic energy $\simeq \Lambda$ their number density is far below the equilibrium value $\sim \Lambda^3$. This is by assumption, since otherwise the glueball sector reheat temperature would be above the confinement scale (as discussed around Eq.~\eqref{eq:tempaboveL}). 
However subsequently $2\rightarrow 3$ processes continue producing new glueballs, reducing the average kinetic energy so the corresponding equilibrium number density, Eq.~\eqref{eq:neq}, drops exponentially.\footnote{The glueball kinetic energy is also redshifted away, however since $3\rightarrow 2$ processes are fast compared to Hubble expansion, $2 \rightarrow 3$ interactions are as well until the chemical equilibrium is reached.} As a result, the system will reach a point where its number density matches the equilibrium value given the average glueball kinetic energy.

Having reached the equilibrium number density, the glueballs evolve as in Section~\ref{sec:high}, with their yield reduced until $3\rightarrow 2$ interactions freeze out. The difference to the thermal scenario (with the hidden sector reheated above its confinement scale) is that the entropy ratio between the hidden and visible sectors is suppressed. Once in chemical equilibrium, the glueball sector entropy is given by non-relativistic formula $s_g \simeq n_g \Lambda/T_g$, where $n_g \simeq 2n_{\rm nr}$, and we define the temperature as $T_g = \Lambda/c$ with $c$ a number that is typically $\sim 10$. Compared to the visible sector entropy this is
\beq \label{eq:rel}
\frac{s_g}{s_v} \simeq  c B \frac{g_g}{g_v^{5/4}}  \frac{m_X^{3/2}}{M_{\rm Pl}^{1/2} \Lambda} ~. 
\eeq
Then the relic density calculation is as in Section~\ref{sec:high}, except with an effective parameter
\beq \label{eq:beff}
B_{\rm eff} = \left(\frac{g_v s_g}{g_g s_v}\right)^{4/3} \simeq \frac{c^{4/3} B^{4/3}}{g_v^{1/3}} \frac{m_{X}^2}{M_{\rm Pl}^{2/3} \Lambda^{4/3}} ~,
\eeq
using Eq.~\eqref{eq:rel}. The suppression of the yield disappears when the glueball sector reheat temperature is $\Lambda$ (given in Eq.~\eqref{eq:tempaboveL}), in which case $B_{\rm eff} = B$ as expected.

The final glueball relic density in this scenario can be straightforwardly calculated using Eq.~\eqref{eq:beff} and the results of Section~\ref{sec:high}. Numerical study shows that the $3\rightarrow 2$ processes have a small effect on the final yield, and typically the effect is $\OO\left(1\right)$ at most compared to the estimate Eq.~\eqref{eq:yapprox3}. The relative importance of late time $3\rightarrow 2$ processes is smaller than in the high temperature reheating case because chemical equilibrium is reached at much lower glueball energies, and $3\rightarrow 2$ processes are only active for a short time.

\bibliographystyle{JHEP}
\bibliography{reference}

\end{document}